\documentclass{ws-mpla}
\usepackage{graphicx}
\usepackage[super,compress]{cite}
\usepackage{tensor}
\usepackage{hyperref}
\usepackage{mathtools}
\usepackage{url}
\usepackage{color}
\usepackage{soul}
\hypersetup{colorlinks,urlcolor=cyan,citecolor=green,linkcolor=blue,filecolor=black}
\begin{document}

\catchline{}{}{}{}{}
\title{Thorne-\.{Z}ytkow objects with mirror neutron star cores?}

\author{Zurab K. Silagadze}
\address{Budker Institute of Nuclear Physics and \\ Novosibirsk State University, 630 090, Novosibirsk, Russia. \\ silagadze@inp.nsk.su}

\maketitle

\pub{Received (Day Month Year)}{Revised (Day Month Year)}
 


 
\begin{abstract}
A Thorne-\.{Z}ytkow object can be formed when a neutron star is absorbed by the envelope of its giant companion star, spirals toward the center of the giant star due to the drag from the surrounding envelope, and merges with the star's core. During in-spiral, dynamical friction is the main drag force acting on a neutron star inside the star envelope. However, dynamical friction is caused by the neutron star's gravitational interaction with its own gravitationally induced wake. Therefore, exactly the same gravitational drag force will act on a mirror neutron star, and if a significant part of the dark matter is in the form of mirror matter, then we expect the formation of Thorne-\.{Z}ytkow objects with a mirror neutron star inside. We explore the plausibility of the existence of such exotic Thorne-\.{Z}ytkow objects.
\end{abstract}

\maketitle

\section{Introduction}
Both in particle physics and cosmology, the mystery of dark matter remains unresolved. Evidence for the existence of dark matter is compelling and comes from different scales \cite{Arbey_2021,Sanders_2010,Buckley_2017,Bertone_2016}. On a galactic scale, flat rotation curves of galaxies are usually interpreted as evidence of the existence of a dark matter halo around these galaxies. At the scale of galaxy clusters, gravitational lensing data, as well as observations of hot gas in clusters, confirm the existence of huge amounts of dark matter in both galaxies and clusters of galaxies. Finally, one more irrefutable evidence of the existence of dark matter was obtained as a result of measurements of the anisotropy of the cosmic microwave background on cosmological scales \cite{Planck_2015}.

Hypothetical mirror matter is the oldest but still viable candidate for dark matter. The idea that every ordinary particle has its own mirror partner was born after the realization that parity nonconservation in weak interactions does not necessarily mean that nature distinguishes between left and right \cite{Lee_1956,Kobzarev_1966,Pavsic_1974}. In the modern language of gauge theories, mirror matter idea was rediscovered in Ref. \refcite{Foot_1991}, and it is based on the gauge symmetry $G_S\times G_M$, where $G_M=G_S=SU(3)\times SU(2)\times U(1)$ is the Standard Model gauge group.

In the minimal case, which we will assume in this article, the particles of the Standard Model are singlets with respect to $G_M$, and vice versa, mirror particles are singlets with respect to $G_S$, with the gravitational interaction being the only common interaction for both the visible and for the hidden (mirror) sector. The fact that the mirror matter could be a hidden sector connected with the visible sector mainly by gravity was realized in Ref. \refcite{Kobzarev_1966}. Thus, Kobzarev, Okun and Pomeranchuk can be considered the founding fathers of mirror dark matter.

In the non-minimal case, the mirror and ordinary (visible) sectors can communicate through mixing of neutral particles. Several such ``portals" have been considered in the literature, including photon portal \cite{Glashow_1986,Foot_2001}, dark photon portal \cite{Alizzi_2021}, neutrino portal \cite{Zel'dovich_1981,Silagadze_1995,Foot_1995,Berezhiani_1995}, neutron portal \cite{Berezhiani_2006,Addazi_2022}, axion portal \cite{Rubakov:1997vp,Berezhiani:2000gh,Giannotti:2005eb}, flavor portal \cite{Berezhiani_flavor1,Berezhiani_flavor2} and Higgs portal \cite{Barbieri_2016,Koren_2020,Harigaya_2020}.

Of course, the idea behind the mirror matter can be generalized by assuming $G_M\ne G_S$, and such a low energy theory with corresponding shadow matter is expected from the heterotic
$E_8\times E_8$ string theory \cite{Kolb:1985bf,Khlopov:2002gg}.

Since the foundations of the theory of mirror matter have been repeatedly expounded in the literature, we will only point out some exemplary reviews \cite{Okun:2006eb,Khlopov:2012lua,Foot:2014mia,Berezhiani:2005ek,Blinnikov:2009nn,Ciarcelluti:2010zz} in which one can find other references on the topic.

In the context of this note, the most important are the astrophysical manifestations of the mirror world discussed in Refs. \refcite{Blinnikov:1982eh,Blinnikov:1983gh,Khlopov:1989fj,Khlopov:2012lua}. In addition to the effects considered in these papers, we point to another astrophysical manifestation of the mirror world: the possible existence of Thorne-\.{Z}ytkow objects with a mirror neutron star inside.

\section{Thorne-\.{Z}ytkow objects}
In 1975 Thorne and \.{Z}ytkow predicted the existence of an entirely new class of stellar objects with massive envelopes and degenerate neutron cores \cite{Thorne_1975}. In fact, already in the 1930s, a number of people, including Gamow \cite{Gamow_1937} and Landau \cite{Landau_1938}, were thinking about stars with a degenerate neutron core. However, a detailed analysis was not carried out and unambiguous conclusions were not drawn \cite{Thorne_1977}. Thorne and \.{Z}ytkow performed \cite{Thorne_1975,Thorne_1977} the first detailed analysis of steady state models of the stellar structure of such stars now called Thorne-\.{Z}ytkow objects (T\.{Z}O). 

It was predicted that  T\.{Z}Os can be formed by three different mechanisms: 
\begin{itemize}
\item The collision of a massive main sequence star and a neutron star (NS) in a globular cluster and the subsequent formation of a stellar binary by tidal capture can, under certain circumstances, lead to the formation of T\.{Z}O \cite{Ray_1987}.
\item The coalescence of a NS with a massive companion in a close binary system \cite{Taam_1978}.
\item A newly formed NS, which received an asymmetric supernova kick in the direction of its companion, is engulfed by this companion \cite{Leonard_1994}.
\end{itemize}
The estimated total birthrate of T\.{Z}Os in the Milky Way galaxy is about $1.5\times 10^{-4}~\mathrm{yr}^{-1}$, while the expected characteristic T\.{Z}O lifetime is $10^5-10^6~\mathrm{yr}$ \cite{Zhu_2018,Rees_1995}. This implies that  there should be about 20-200 T\.{Z}Os in the Galaxy at present \cite{Rees_1995}.

Recently, a new channel for the formation of T\.{Z}Os has been proposed, which may make an important contribution to the T\.{Z}Os population \cite{Ablimit_2022}.

Since the vast red giant envelope conceals the conditions of these objects' inner regions from a remote observer, it will be challenging to find T\.{Z}Os through astrophysical observations. A  possible observational hallmark of such stars, according to Thorne and \.{Z}ytkow, may be chemical abnormalities in their atmospheres.

To a distant observer, T\.{Z}O looks like a red giant or red supergiant with a luminosity $10^4-10^5$ higher than that of the Sun, with a surface temperature of about 3000 K, and a photosphere radius of about $1000~\mathrm{R}_\odot$. Large diffuse envelope of T\.{Z}O is separated from its core by a very thin (from few tens to few hundred meters) energy-generation layer. For low envelope mass, the main energy source is gravitational energy released by the contracting matter \cite{Thorne_1977,Cannon_1992}. When the envelope mass exceeds approximately $10~\mathrm{M}_\odot$, a hydrogen burning shell appears at the bottom of the envelope with a temperature of up to $10^9~K$, and hydrogen burning becomes the most important source of energy \cite{Thorne_1977,Cannon_1992}.

Nucleosynthesis in T\.{Z}O differs significantly from that in normal (super)giants. The primary energy sources in massive ($>1.3~\mathrm{M}_\odot$) stars during their hydrogen burning phase are the so-called CNO cycles. The CNO cycle is a chain of nuclear reactions with $\prescript{12}{}{\mathrm{C}}$ or $\prescript{16}{}{\mathrm{O}}$ as the catalytic material, facilitating the efficient fusing of four hydrogen nuclei into one helium nucleus \cite{Wiescher_2010,Wiescher_2018}. The example of CNO cycle is
\begin{align}
&  \prescript{12}{}{\mathrm{C}}+\prescript{1}{}{\mathrm{H}}\to \prescript{13}{}{\mathrm{N}}+\gamma,\;\;\prescript{13}{}{\mathrm{N}}\to \prescript{13}{}{\mathrm{C}}+e^++\nu_e,\;\;\prescript{13}{}{\mathrm{C}}+\prescript{1}{}{\mathrm{H}}\to \prescript{14}{}{\mathrm{N}}+\gamma,\;\; \nonumber \\
& \prescript{14}{}{\mathrm{N}}+\prescript{1}{}{\mathrm{H}}\to \prescript{15}{}{\mathrm{O}}+\gamma,\;\;\prescript{15}{}{\mathrm{O}}\to \prescript{15}{}{\mathrm{N}}+e^++\nu_e,\;\; \prescript{15}{}{\mathrm{N}}+\prescript{1}{}{\mathrm{H}}\to \prescript{12}{}{\mathrm{C}}+\prescript{4}{}{\mathrm{He}}.
\nonumber
\end{align}
In T\.{Z}O, at the bottom of the convective envelope where the combustion takes place, material circulates in a time of about $0.01~\mathrm{s}$. The time scales of some $\beta^+$-decays from CNO-cycles are much larger, up to $100~\mathrm{s}$. Therefore, the chain of reactions hangs in anticipation of $\beta^+$-decays \cite{Cannon_1993}.

At a very high temperature in the T\.{Z}O burning zone, an alternative chain of reactions involving the rapid capture of protons on seed nuclei (rp-process) takes place \cite{Wallace_1981}. Convection transports the seed nucleus (originally C, N, or O) into the combustion zone, where it burns rapidly through the repeated addition of protons to a proton-rich species. The rp-process stops when the proton addition time exceeds the turbulent turnover time. Then the end product of the rp-process is returned to the envelope by convection, where it wanders randomly until it undergoes $\beta^+$-decay. If the decay occurs near the combustion zone, it is likely that the resulting nucleus will be transferred back to the combustion zone and burn again. If the $\beta^+$-decay time is long, then the transformed seed will probably be transferred to the star's photosphere and one will be able to observe its final decay products (stable or long-lived isotopes) \cite{Biehle_1981,Biehle_1994}.

Because nuclear reactions in the T\.{Z}O envelope proceed in a highly unconventional manner, T\.{Z}O is expected to have a peculiar surface abundances of elements heavier than iron \cite{Biehle_1994,Cannon_1993}, and this is considered as the observational signature of T\.{Z}O.
The interrupted rapid proton process (rapid proton process interrupted by convection) can lead to excess production of heavy elements such as rubidium, strontium, yttrium and molybdenum. T\.{Z}Os are also expected to be enriched in $\prescript{40}{}{\mathrm{Ca}}$ and  $\prescript{7}{}{\mathrm{Li}}$. However, the nucleosynthetic signal is not unique, as there are several T\.{Z}O imposter stars that mimic the T\.{Z}O nucleosynthetic signal \cite{Farmer_2023}. A number of potential T\.{Z}O candidates have been identified, most notably HV 2112 \cite{Levesque_2014}, but not without controversy \cite{Farmer_2023,Beasor_2018}.

HV 2112 is the most exceptional red supergiant star in the Small Magellanic Cloud. Many of its properties are consistent with those expected for T\.{Z}O. However, the same spectroscopic features are expected for super-asymptotic giant branch (sAGB) stars too, except that HV 2112 shows a slight increase in $\mathrm{Ca}$ compared to $\mathrm{K}$, which is not expected for sAGB stars. On the other hand, there are three possible obstacles to identifying HV 2112 as T\.{Z}O: its estimated mass is slightly less than the minimum stable mass of T\.{Z}O from the model in Ref. \refcite{Cannon_1993}, the observed $\mathrm{Ca}$ enhancement cannot be produced in T\.{Z}O (but it may be produced in the core of the progenitor star and may be mixed upward towards the photosphere during formation of the T\.{Z}O), finally, some  U-shaped structure is observed around HV 2112, which may be a wind bubble inflated by the progenitor of the sAGB star (but it may also be a fluctuation of the interstellar medium not associated with HV 2112) \cite{O‘Grady_2023,Tout_2014}. 

At present, the true nature of HV 2112 cannot be unambiguously established. Both further observations and modeling are needed to clarify the situation. Note that the T\.{Z}O models have not been revised since their debut several decades ago, and many of the theoretical predictions of the observed T\.{Z}Os spectra were made before significant changes in stellar evolution codes \cite{O‘Grady_2023}.

\section{Dynamical friction in a supergiant's envelope}
The NS in-spiral phase is an important part of the T\.{Z}O formation process. The common envelope evolution during in-spiral is still poorly understood, since the wide range of spatial and temporal scales involved and the lack of symmetry pose a serious problem \cite{Ropke_2023}.

As the NS enters the envelope, various drag forces act on it including hydrodynamic friction (ram pressure), gravitational drag due to Bondi–Hoyle–Lyttleton accretion, gravitational drag due to deflection of the surrounding gas imparting momentum to the NS, and tidal forces \cite{Ropke_2023}. In fact, hydrodynamic friction is completely negligible compared to gravitational effects, as we will demonstrate below.

\begin{figure}[htp]
\includegraphics[scale=0.7]{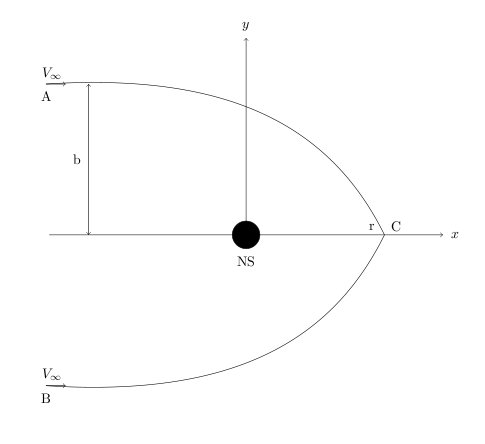}
\caption{Schematic geometry illustrating the Hoyle-Lyttleton accretion idea.}
\label{fig}
\end{figure}

In his classic work \cite{Chandrasekhar_1943} Chandrasekhar came to the conclusion about the existence of dynamical friction (gravitational drag) from very general considerations. In our case, the physical picture is quite simple: in the gravitational wake of a moving NS, more mass is located downstream than upstream, and this causes a braking gravitational attraction of the NS. Let us estimate the resulting drag force using the approach of Hoyle and Lyttleton \cite{hoyle_lyttleton_1939} illustrated by Fig.\ref{fig}.

In the rest frame of the NS, the cloud of envelope particles moves from left to right with a velocity at a large distance from the NS equal to $V_\infty$ (the velocity of the NS relative the envelope). Consider a streamline $AC$ with impact parameter $b$. Particles of this streamline will collide at point $C$ with particles of the opposite streamline $BC$ and, as a result, will lose the vertical component ($y$-component) of their velocity $V_y$. At that, if the remaining $x$-component of their velocity $V_x$ at this point is less than the escape velocity, then the particles will fall and accrete onto the NS.

To find the critical impact parameter below which the accretion begins, it is useful to use the Hamilton vector \cite{Martinez-y-Romero_1993,Gonzalez-Villanueva_1996}, ``the lost sparkling diamond of introductory level mechanics" \cite{Chashchina_2008,Shatilov_2021}. The Hamilton vector is a vector constant of motion in the Kepler problem and has the form
\begin{equation}
\vec{h}=\vec{V}-\frac{\alpha}{L}\,\vec{e}_\varphi,
\label{eq1}
\end{equation}
where $\alpha=G_NmM_{NS}$ is the gravitational coupling constant for the streamline particle with mass $m$, $M_{NS}$ is the NS mass, $L=mV_\infty b$ is the particle's angular momentum, and $\vec{e}_\varphi$ is the unit vector of the polar coordinate system in the $\varphi$-direction. The great pedagogical advantage of the Hamilton vector is that it can be introduced in a simple and natural way, unlike its more famous relative, the Laplace-Runge-Lenz vector \cite{Shatilov_2021}.

Note that very far from the origin of coordinates (at the origin the NS is located) $\vec{e}_\varphi$ is perpendicular to the $x$ axis and directed downward, while at point $C$ $\vec{e}_\varphi$ is still perpendicular to the $x$ axis, but directed upward. Therefore,
\begin{equation}
(h_{\infty})_x=V_\infty,\;\; (h_{\infty})_y=\frac{\alpha}{L},\;\;\;\;  
(h_{C})_x=(V_{C})_x,\;\; (h_{C})_y=(V_{C})_y-\frac{\alpha}{L}.
\label{eq2}
\end{equation}
The conservation of the vector $\vec{h}$ implies
\begin{equation}
(V_C)_x=V_\infty,\;\; (V_C)_y=\frac{2\alpha}{L}.
\label{eq3}
\end{equation}
The condition for accretion thus is
\begin{equation}
\frac{m(V_c)_x^2}{2}-\frac{\alpha}{r}=\frac{mV_\infty^2}{2}-\frac{\alpha}{r}<0.
\label{eq4}
\end{equation}
On the other hand, conservation of energy gives
\begin{equation}
\frac{mV_\infty^2}{2}=\frac{m}{2}\left [ (V_c)_x^2+(V_c)_y^2\right ]-\frac{\alpha}{r}=
\frac{mV_\infty^2}{2}+\frac{2m\alpha^2}{L^2}-\frac{\alpha}{r},
\label{eq5}
\end{equation}
which imply
\begin{equation}
\frac{\alpha}{r}=\frac{2m\alpha^2}{L^2}.
\label{eq6}
\end{equation}
We can substitute this result for $\alpha/r$ into (\ref{eq4}), and since $L=mV_\infty b$ the accretion condition becomes
\begin{equation}
b<\frac{2G_NM_{NS}}{V_\infty^2}=R_{HL}.
\label{eq7}
\end{equation}
Particles with an impact parameter smaller than the critical value $R_{HL}$, known as the Hoyle–Lyttleton radius, will be accreted. The Hoyle–Lyttleton accretion rate is therefore
\begin{equation}
\dot{M}_{HL}=\pi R_{HL}^2V_\infty\rho_\infty=\frac{4\pi\rho_\infty G_N^2 M_{NS}^2}{V_\infty^3},
\label{eq8}
\end{equation}
where $\rho_\infty$ is the density of the unperturbed envelope (far from the NS).

For our purposes, this simplified interpretation of accretion is sufficient. Subsequent developments following the pioneering work of Hoyle and Littleton are described in the pedagogical review of Bondi-Hoyle-Littleton accretion \cite{Edgar_2004}, where in particular (\ref{eq8}) is obtained by a different method.

The drag force due to Hoyle-Littleton accretion can be estimated as
\begin{equation}
F_{HL}=\dot{M}_{HL}V_\infty=\frac{4\pi\rho_\infty G_N^2 M_{NS}^2}{V_\infty^2}.
\label{eq9}   
\end{equation}

More sophisticated treatment \cite{Chandrasekhar_1943,Binney_2008,Ostriker:1998,Tremaine_1984} leads essentially to the conclusion that the Hoyle-Littleton result should be multiplied by the so-called Coulomb logarithm $\Lambda$:
\begin{equation}
F_{dyn}=\frac{4\pi\rho_\infty G_N^2 M_{NS}^2}{V_\infty^2}\,\Lambda,
\label{eq10}   
\end{equation}
where in the supersonic case \cite{Ostriker:1998}
\begin{equation}
\Lambda\approx \frac{1}{2}\ln{\left(1-\frac{c_s^2}{V^2}\right)}+\ln{\left(\frac{b_{max}}{b_{min}}\right)},
\label{eq11}
\end{equation}
with $c_s$ being the speed of sound in the envelope (so  $V/c_s$ is the Mach number). 
The parameters $b_{max}$ and $b_{min}$, which are difficult to determine precisely, represent the maximum and minimum impact parameters contributing to the drag. For estimation purposes $b_{max}$ can be assumed to be of the order of the envelope size and $b_{ min}\approx \mathrm{max}(G_NM_{NS}/V_\infty^2,R_{NS})$ \cite{Tremaine_1984}.

Hydrodynamical drag force is proportional to the surface area of the NS and can be expressed in the form 
\begin{equation}
F_{hyd}=\frac{1}{2}C_d\rho_\infty V_\infty^2\pi R_{NS}^2,   
\label{eq12}
\end{equation}
where $C_d\approx 0.9$ is  the  dimensionless  drag  coefficient  for  a sphere \cite{Villaver_2009}.

Equations (\ref{eq10}) and (\ref{eq12}) imply
\begin{equation}
\frac{F_{hyd}}{F_{dyn}}=\frac{C_d}{8\Lambda}\,\frac{R_{NS}^2V_\infty^4}{G_N^2M_{NS}^2}\approx
10^{-11}\,\frac{C_d}{\Lambda}.
\label{eq13} 
\end{equation}
It is clear that the hydrodynamical drag is completely negligible compared to the dynamical friction (gravitational drag). In (\ref{eq13}) we have used $R_{ns}\approx 10~\mathrm{km}$, $M_{NS}\approx 2M_\odot$, and $V_\infty\approx 500~\mathrm{km}/\mathrm{s}$.

\section{Binary systems of mixed mirrority}
As we see, after NS enters the companion’s envelope, its fate is determined solely by gravity. Therefore, nothing will change significantly if the NS is replaced by a mirror NS, and under some circumstances the formation of Thorne-\.{Z}ytkow objects with mirror neutron stars inside is expected.

Although the visible and mirror sectors have the same microphysics, no symmetry between the two sectors is possible at the macroscopic level without directly contradicting, for example, the Big Bang nucleosynthesis bounds on the effective number of light neutrinos.
A special inflationary scenario may lead to a difference in the initial temperatures of the two sectors, with a lower temperature of the mirror sector and avoid such bounds \cite{Berezhiani:2000gw}.

It is expected that the macroevolution of the mirror sector should deviate significantly from standard cosmology with respect to such crucial epochs as baryogenesis, nucleosynthesis, etc. In particular, as a result of the lower initial temperature, the mirror sector is expected to produce greater baryon asymmetry than the visible sector, mirror stars will be dominated by helium, they will evolve faster than normal stars, and massive mirror stars will explode as mirror supernovae leaving behind mirror neutron stars or black holes \cite {Berezhiani:2005vv}.

Given mirror neutron stars, the crucial question is whether T\.{Z}O formation scenarios apply to them. The first scenario for the formation of T\.{Z}O, as mentioned above, assumes the formation of a binary system through tidal capture during the collision of a massive main sequence star and a neutron star in a globular cluster. 

Even initially homogeneously mixed ordinary and mirror baryonic matter will separate to form objects of a definite mirrority  up to the scale of globular clusters, on scales at which thermal instability plays a significant role \cite{Khlopov:2012lua}. This occurs due to the separate development of thermal instability in ordinary and mirror matter. However, a globular cluster has several opportunities to capture stars of a different mirrority. For example, this can happen when the formation of a globular cluster occurs in gas captured by a globular cluster of opposite mirrority during the separation of ordinary and mirror matter \cite{Khlopov:1989fj,Khlopov:2012lua}. Therefore, there will be a possibility that a mirror neutron star will collide with an ordinary massive star in such a globular cluster, and the first scenario is expected to work in the case of a mirror neutron star T\.{Z}O too, although less efficiently than in the ordinary T\.{Z}0 case.

The second scenario for the formation of T\.{Z}O assumes the coalescence of a NS with a massive companion in a close binary system. A binary system with stars of different mirrority can be formed when a globular cluster captures a star of opposite mirrority. However, much more efficient source of such binaries are giant molecular clouds. 

Giant molecular clouds have a mass comparable to a globular cluster, but are an order of magnitude more numerous than globular clusters in the Galaxy. Therefore, it is expected that the interpenetration of ordinary and mirror molecular clouds into each other will occur quite often, and since the clouds collide with low relative velocities and are highly dissipative systems, in some cases giant molecular clouds of mixed mirrority can form. The  star  formation process in such clouds can produce binaries of different mirrority with enhanced probability \cite{Khlopov:1989fj,Khlopov:2012lua}. Subsequent evolution of such a mixed close binary system can lead to the formation of a NS, which can remain bound to its companion of different mirrority. 

Since the formation of binary systems of mixed mirrority is possible, the third scenario for the formation of T\.{Z}O is also valid: NS receives a kick during an asymmetric supernova explosion, which sends it to its companion of the opposite mirrority and as a result it becomes engulfed by the companion’s envelope.

The probability of formation of the mixed (ordinary+mirror) star binaries is expected to be very low \cite{Blinnikov:2009nn}. Therefore, to somewhat compensate this small probability, it is desirable to have much more mirror neutron stars than the ordinary ones, and more generally, much more mirror matter than ordinary matter. We know that the universe contains five times more dark matter than the ordinary matter. However, is it possible that all dark matter is in the form of mirror matter?

In the case of a perfect out-of-equilibrium condition in both sectors, one would expect the generated baryon asymmetries $\eta = n_B/n_\gamma$ and $\eta^\prime = n^\prime_B/n^\prime_\gamma$ in two sectors will not differ much, $\eta^\prime\approx\eta$, if baryon asymmetry in both sectors is generated by the same mechanisms of baryogenesis. On the other hand, photon densities in two sectors are related according to $n^\prime_\gamma/n_\gamma=x^3$, there $x=T^\prime/T$ is the ratio of temperatures of the mirror and ordinary sectors. Therefore, since mirror particles have exactly the same masses as ordinary particles, we get \cite{Berezhiani:2000gw}
\begin{equation}
\beta=\frac{\Omega^\prime_B}{\Omega_B}=\frac{n^\prime_B}{n_B}=\frac{\eta^\prime}{\eta}\,\frac{n^\prime_\gamma}{n_\gamma}\approx x^3,
\label{eq_add}
\end{equation}
and we are in trouble, since the cosmological limits imposed by the large-scale structure and anisotropy of the cosmic microwave background imply a rather strong bound $x < 0.2-0.3$, as shown in Ref. \refcite{Berezhiani_LSS}. Therefore $\beta\ll 1$ and mirror dark matter becomes phenomenologically uninteresting in astrophysics.

But there is a way out. One can imagine a mechanism that allows large baryon asymmetry in the mirror sector without contradicting restrictions on the temperature ratio. In Refs.\cite{Bento_Lepto1,Bento_Lepto2} a leptogenesis scenario was suggested for a unified picture for the baryon and mirror baryon asymmetries. In both sectors, baryon asymmetry arises due to out-of-equilibrium, $B-L$ and $CP$-violating scattering processes mediated, for example, by heavy Majorana neutrinos that transform ordinary particles into mirror particles, and in particular lead to neutrino - mirror neutrino oscillations. The small factor $x^3$ disappears altogether. Instead, the difference between $\Omega^\prime_B$ and $\Omega_B$ is due to the different effectiveness of the various washout effects: they are expected to be more effective in the hotter ordinary sector than in the cooler mirror sector. It has been shown \cite{Berezhiani_Lepto} that $\beta\sim 5$ can be obtained in this way without fine-tuning.
 
A rough estimate of the abundance of mirror neutron stars in the galaxy can come from the rate of the gravitational wave signals from `invisible' mergers in the mirror NS binaries. It was argued that only one in ten binary neutron star or neutron star-black hole mergers detected by LIGO/Virgo would be accompanied by a gamma-ray burst and other electromagnetic signals if mirror particles constitute the bulk of the universe's dark matter \cite{Beradze_2019}. Such kind of observations and theoretical considerations do not exclude that in the mirror world, mergers of massive first-generation stars are more probable, and compact second-generation objects, such as neutron stars, form at a higher rate \cite{Beradze_2021}.

\section{Conclusions}
As we see, gravity is the main actor at the stage of formation of T\.{Z}O. Therefore, if mirror matter really exists and constitutes a significant, if not dominant, part of dark matter, then the formation of Thorne-\.{Z}ytkow objects with the cores from mirror neutron stars is expected. Of course, a mirrored situation is also possible: Thorne-\.{Z}ytkov objects with  ordinary neutron star cores and mirror matter envelope.

Such exotic objects can be very rare, since, in addition to gravity, their formation requires encounters or the formation of binary systems of stars with different mirrority, and ordinary and mirror matter are well separated on the stellar scale \cite{Blinnikov:2009nn}. Nevertheless, the formation of mixed mirrority systems are possible in variety of ways, briefly mentioned above and described in more detail in Ref. \refcite{Khlopov:2012lua}.

Steady-state models of exotic T\.{Z}O with mirror matter require special attention, since ordinary matter in free fall will not stop at the surface of the mirror neutron star core, but will continue to fall towards the center, where something like an inertial thermonuclear reactor will be realized.

Since in the case of exotic T\.{Z}O with mirror matter the depth of the potential well into which the matter falls is about 1.5 times greater than in the case of ordinary T\.{Z}O, we expect that the isotopic fingerprint of such T\.{Z}O will be different, and this can be used as an observational signature of such objects. Indeed, If some small piece of ordinary matter far from the core in the envelope begins free fall towards the core, then in the case of an ordinary NS it will stop at the surface of the NS, and in the case of a mirror NS it will continue to free fall towards the center of the NS and, thus, gain about 1.5 times more energy reaching the center, compared to the first case. Therefore, the temperature of ordinary matter during nucleosynthesis reactions near the center of the mirror NS will be higher than the corresponding temperature in the combustion zone of ordinary T\.{Z}O. Specific predictions of expected isotopic anomalies under such extreme conditions require careful modeling and are beyond the scope of this paper.

Combined with observations of unusual surface chemistry, detection of continuous gravitational waves from the rotating neutron star core, which should be possible with Advanced LIGO and future gravitational wave detectors, could provide compelling evidence for the existence of Thorne-\.{Z}ytkow objects \cite{DeMarchi:2021vwr}. Of course, the gravitational wave signal by itself cannot distinguish ordinary T\.{Z}O from the exotic one, unless the unusual steady-state structure of the exotic T\.{Z}O affects the properties of the gravitational wave signal.

One can  expect also coalescences of ordinary NS with the mirror star. In this case it is possible that T\.{Z}O, with  ordinary neutron star core and mirror matter envelope, may look like an unusual neutron star when observed due to the presence of mirror matter inside the neutron star \cite{Sandin:2008db}, especially if there exits some portals connecting two worlds \cite{Berezhiani_MNS}.

The mirror matter model is well motivated by the desire to restore the symmetry of nature relative to spatial inversion. However, the Universe is a strange place \cite{Wilczek:2005ez}, and it is not excluded that $Z_2$-symmetry implied by the mirror matter model is spontaneously broken \cite{Berezhiani_1995,Berezhiani_Dolgov,Berezhiani_brokenP,Hippert:2021fch}. The possibility of spontaneously broken mirror parity and the nature of mirror matter and mirror stars in this case were first discussed in \cite{Berezhiani_Dolgov,Berezhiani_brokenP}.
Another possibility is the possible existence of various portals between the visible and mirror worlds mentioned in the introduction. A discussion of T\.{Z}Os in these more general situations is beyond the scope of this article. We just mention an interesting possibility that in case of the neutron-mirror neutron portal the creation of ordinary matter (or antimatter) ``egg" is expected inside the mirror NS \cite{Berezhiani_MNS,Berezhiani_atimatter}. The case of the antimatter ``egg" is especially interesting because some part of this antimatter can be released in space if the mirror NS receives a kick in some catastrophic event, thus explaining the presence of antinuclei in cosmic rays if they will be found by AMS-02 experiment \cite{Berezhiani_atimatter}.

We are accustomed to the fact that many of the symmetries of the Standard Model and its proposed extensions are broken. Therefore, from a modern point of view, the exact parity symmetry cannot be considered as a strong basis for the idea of mirror matter. However, after more than 60 years of studying mirror matter models, a consistent and coherent picture has emerged, mainly due to the works of Z. Berezhiani, R. Foot and their collaborators, which makes the idea of mirror matter plausible and useful for explaining some of the mysteries of nature,  like resolving the anomalies of the flavor symmetries and understanding the origin of the suppression of the flavor-changing processes \cite{Berezhiani_flavor1}.

\section*{Aknowledgements}
We are grateful to the reviewers for useful comments.

\bibliography{Mirror_TZ}

\end{document}